\documentclass[twocolumn,english,pra]{revtex4}
\usepackage[T1]{fontenc}
\usepackage[latin9]{inputenc}
\usepackage{amsmath}
\usepackage{amssymb}
\usepackage{graphicx}
\usepackage{esint}

\makeatletter
\@ifundefined{textcolor}{}
{%
 \definecolor{BLACK}{gray}{0}
 \definecolor{WHITE}{gray}{1}
 \definecolor{RED}{rgb}{1,0,0}
 \definecolor{GREEN}{rgb}{0,1,0}
 \definecolor{BLUE}{rgb}{0,0,1}
 \definecolor{CYAN}{cmyk}{1,0,0,0}
 \definecolor{MAGENTA}{cmyk}{0,1,0,0}
 \definecolor{YELLOW}{cmyk}{0,0,1,0}
 }

\makeatother

\usepackage{babel}
\begin{document}

\title{High temperature thermodynamics of strongly interacting $s$-wave
and $p$-wave Fermi gases in a harmonic trap}

\author{Shi-Guo Peng$^{1,2}$, Shi-Qun Li$^{1}$, Peter D. Drummond$^{2}$
and Xia-Ji Liu$^{2}$}

\affiliation{$^{1}$Department of Physics, Tsinghua University, Beijing, 100084,
China}

\affiliation{$^{2}$Centre for Atom Optics and Ultrafast Spectroscopy, Swinburne
University of Technology, Melbourne 3122, Australia}

\date{\today{}}
\begin{abstract}
We theoretically investigate the high-temperature thermodynamics of
a strongly interacting trapped Fermi gas near either $s$-wave or
$p$-wave Feshbach resonances, using a second order quantum virial
expansion. The second virial coefficient is calculated based on the
energy spectrum of two interacting fermions in a harmonic trap. We
consider both isotropic and anisotropic harmonic potentials. For the
two-fermion interaction, either $s$-wave or $p$-wave, we use a pseudopotential
parametrized by a scattering length and an effective range. This turns
out to be the simplest way of encoding the energy dependence of the
low-energy scattering amplitude or phase shift. This treatment of
the pseudopotential can be easily generalized to higher partial-wave
interactions. We discuss how the second virial coefficient and thermodynamics
are affected by the existence of these finite-range interaction effects.
The virial expansion result for a strongly interacting $s$-wave Fermi
gas has already been proved very useful. In the case of $p$-wave
interactions, our results for the high-temperature equation of state
are applicable to future high-precision thermodynamic measurements
for a spin-polarized Fermi gas near a $p$-wave Feshbach resonance.
\end{abstract}
\maketitle

\section{Introduction}

Strongly interacting fermions occur in many fields of physics, ranging
from condensed matter physics to nuclear physics, astrophysics and
cosmology. Recently, a new type of fermionic superfluid has been realized
in ultracold atomic gases of $^{40}$K and $^{6}$Li confined in a
harmonic trap. In these systems the coupling strength between the
fermions can be tuned precisely by an external magnetic field from
weak to infinitely strong across a collisional (Feshbach) resonance.
In the case of $s$-wave interactions, this has already led to the
observation of a smooth crossover from a Bardeen-Cooper-Schrieffer
(BCS) superfluid to a Bose-Einstein condensation (BEC). By contrast,
in a Fermi system with $p$-wave interactions, a quantum phase transition
is anticipated to take place instead of a smooth crossover\cite{SadeMelo,SadeMelo2005,Gurarie}. 

By sweeping the magnetic field, $p$-wave Feshbach resonances have
recently been observed by several experimental groups \cite{ENS,Jin,Esslinger,MIT,SUT,Zimmermann}.
The binding energies, lifetimes and radio-frequency spectroscopy of
$p$-wave Feshbach molecules and some low-dimensional aspects of $p$-wave
interactions have been investigated in these experiments. It is well
known that in the dilute, low-energy limit, a two-particle $s$-wave
interaction can be well characterized using a single parameter, the
scattering length. However, in $p$-wave systems a new length scale
- the effective range of interactions - is required, in addition to
the scattering volume \cite{Idziaszek,SadeMelo}. To avoid complicated
calculations in many-body systems with $p$-wave interactions, some
theoretical descriptions \cite{Machida,Machida2010} only use a constant
scattering volume. This is based on the idea of a pseudopotential,
which was first introduced by Huang and Yang \cite{Huang1957}. Recently,
there have been several developments of the pseudopotential method
for higher-partial wave interactions \cite{Stampfer2008,Reichenbach,Pricoupenko,Blume,Stock,Derevianko,Bolda2002,Roth,Idziaszek},
using an energy-dependent scattering length. In particular, numerically
exact solutions of two interacting fermions in a harmonic trap have
been studied with either $s$-wave or $p$-wave couplings.

Given the interesting and unusual nature of $p$-wave interactions,
in this paper we aim to investigate the high-temperature thermodynamics
of a many-body, strongly interacting $p$-wave Fermi gas in a harmonic
trap. The $s$-wave high-temperature thermodynamics will be also included
for completeness. Our investigation is based on the virial expansion
of the thermodynamic potential, which provides a natural bridge between
few-body and many-body physics \cite{HO,Liu2009,HU2010,HU2010PRA,Ho2004}.
Restricting ourselves to the leading order of the expansion, we calculate
the second virial coefficient from two-fermion solutions. Consequently,
we obtain the high-temperature equation of state in the strongly interacting
regime, giving the energy and entropy as a function of temperature.
We address in particular the role of the effective range of $p$-wave
interactions in the thermodynamic state equation.

The calculation of thermodynamics in strongly interacting quantum
many-body systems is always a challenge \cite{HoUniversality,HLDUniversality,HLDTmatrixTheory1,HLDTmatrixTheory2}.
The only conclusive method seems to be that of $ab-initio$ quantum
Monte Carlo simulation. This, however, often suffers from the sign
problem for fermions. Our virial expansion approach to thermodynamics
can provide an accurate benchmark for these simulations at high temperatures.
Already, for $s$-wave interactions, the virial expansion has already
been shown to be very useful in understanding recent thermodynamic
measurements. We anticipate that our $p$-wave virial prediction will
also give valuable insights and possible calibration procedures for
future experiments on a Fermi gas near a $p$-wave resonance.

The calculation of the second virial coefficient requires the whole
energy spectrum of the two-fermion bound states. To facilitate this
calculation, we characterize the pseudopotential using two parameters,
the scattering length and the effective range. This gives a simplified
description for an energy-dependent scattering length. We validiate
our two-parameter pseudopotential treatment by comparing the resulting
spectrum with the spectrum of the full energy-dependent pseudopotential.

This paper is organized as follows. In the next section, we briefly
review the derivation of the pseudopotential method for all partial-wave
interactions and then introduce the two-parameter pseudopotential.
The necessity of including a finite range for $p$-wave interactions
is emphasized. In Sec. III, we present exact solutions for the energy
spectrum of two-fermion systems with either $s$-wave or $p$-wave
interactions in a harmonic trap. In Sec. IV, we calculate the second
virial coefficients. Then, in Sec. V we investigate the high-temperature
thermodynamics of a strongly correlated trapped Fermi gas. Finally,
Sec. VI is devoted to conclusions and final remarks.

\section{Two-parameter pseudopotential}

The first step in understanding the sophisticated physics of a quantum
many-body system is to model the fundamental two-body interactions.
At sufficiently low temperatures the Heisenberg uncertainty principal
means that particles must \emph{spread} over a distance much larger
than the range of the interaction potential. In this case the details
of the true interparticle potential become irrelevant and it may be
modeled by a pseudopotential. The basic idea of the pseudopotential
is to replace the real interaction potential by zero-range potential,
which acts only at $\mathbf{r}=0$ and reproduce the same asymptotic
behavior of the wave function as would occur with the real potential.
The first attempt to derive a generalized pseudopotential was made
by Huang and Yang \cite{Huang1957}. This has been improved by a number
of authors. In this section we briefly review the derivation of the
pseudopotential method \cite{Idziaszek,Stampfer2008}.

\subsection{Two-particle scattering}

We consider a two-particle scattering process and assume that $V(\mathbf{r})$
is the interaction potential. The motion of the center-of-mass can
be separated from the relative motion, and the Schrödinger equation
in the relative coordinate can be written as 
\begin{equation}
\frac{\hbar^{2}}{2\mu}\left(\nabla^{2}+k^{2}\right)\psi\left(\mathbf{r}\right)=V(\mathbf{r})\psi\left(\mathbf{r}\right),\label{eq:pseu1}
\end{equation}
where $k^{2}=2\mu E/\hbar^{2}$ and $\mu=m/2$ is the reduced mass,
assuming identical particles apart from spin. The wavefunction has
the following asymptotic behavior outside the potential range,
\begin{equation}
\psi_{a}\left(\mathbf{r}\right)=\sum_{l=0}^{\infty}\sum_{m=-l}^{l}C_{lm}\left[j_{l}(kr)-\tan\delta_{l}n_{l}(kr)\right]Y_{lm}(\theta,\varphi).\label{eq:pseu2}
\end{equation}
Here $Y_{lm}(\theta,\varphi)$ are spherical harmonics, $j_{l}(kr)$
and $n_{l}(kr)$ are spherical Bessel and Neumann functions respectively,
and $\delta_{l}$ is the phase shift of $l$-th partial wave determined
by the boundary condition. 

Following Huang and Yang \cite{Huang1957}, we extend $R_{l}\left(r\right)=j_{l}(kr)-\tan\delta_{l}n_{l}(kr)$
to the vicinity of the origin at $\mathbf{r}=0$. The real interaction
potential $V(\mathbf{r})$ is now replaced by the pseudopotential
$V_{ps}\left(\mathbf{r}\right)$. We have
\begin{equation}
V_{ps}\left(\mathbf{r}\right)\psi\left(\mathbf{r}\right)=\frac{\hbar^{2}}{2\mu}\left(\nabla^{2}+k^{2}\right)\psi_{a}\left(\mathbf{r}\right)\mid_{r\rightarrow0}.\label{eq:pseu3}
\end{equation}
The asymptotic functions of $j_{l}(kr)$ and $n_{l}(kr)$ at $r\rightarrow0$
are given by,

\begin{gather*}
j_{l}\left(kr\right)\simeq\frac{\left(kr\right)^{l}}{\left(2l+1\right)!!}
\end{gather*}
and

\begin{gather*}
n_{l}(kr)\simeq-\frac{(2l-1)!!}{(kr)^{l+1}},
\end{gather*}
respectively. Thus, only $n_{l}(kr)$ is singular at small $r$. To
solve Eq.(\ref{eq:pseu3}), we use the method introduced in Idziaszek's
paper \cite{Idziaszek} and obtain the pseudopotential for all partial
waves as
\begin{eqnarray}
V_{ps}\psi\left(\mathbf{r}\right) & = & \sum_{l=0}^{\infty}\sum_{m=-l}^{l}\frac{(-)^{l}16\pi^{2}}{(2l+1)!}g_{l}Y_{lm}(\partial)\delta(\mathbf{r})\nonumber \\
 &  & \times\left[\frac{\partial^{2l+1}}{\partial r^{\prime2l+1}}r^{\prime2l+1}Y_{lm}^{*}(\partial^{\prime})\psi\left(\mathbf{r}'\right)\right]_{r^{\prime}=0},\label{eq:pseu5}
\end{eqnarray}
where 
\begin{equation}
g_{l}=-\frac{\hbar^{2}}{2\mu}\frac{\tan\delta_{l}}{k^{2l+1}}.\label{eq:pseu6}
\end{equation}
The partial differential operator $Y_{lm}(\partial)$ is obtained
from the harmonic polynomial $r^{l}Y_{lm}(\hat{r})$ by replacing
the Cartesian coordinates $x_{k}$ with the partial derivatives $\partial_{x_{k}}$
\cite{MacRobert}. According to Eq.(\ref{eq:pseu5}), we can easily
obtain the form of pseudopotential for any specific partial wave.
Using the expression 
\begin{equation}
\frac{4\pi}{2l+1}\sum_{m=-l}^{l}Y_{lm}(\partial)Y_{lm}^{*}(\partial^{\prime})=\sum_{k=0}^{\left[l/2\right]}c_{k}\left(\mathbf{\boldsymbol{\nabla}}\cdot\mathbf{\boldsymbol{\nabla^{\prime}}}\right)^{l-2k}\nabla^{2k}\nabla^{\prime2k}\label{eq:pseu9}
\end{equation}
with $c_{k}=(-)^{k}(2l-2k)!/\left(2^{l}k!(l-k)!(l-2k)!\right)$, we
can write the pseudopotential for $s$- and $p$-wave as

\begin{equation}
V_{s}=-\frac{2\pi\hbar^{2}}{\mu}\frac{\tan\delta_{0}}{k}\delta(\mathbf{r})\frac{\partial}{\partial r}r\label{eq:pseu7}
\end{equation}
and 
\begin{equation}
V_{p}=-\frac{\pi\hbar^{2}}{\mu}\frac{\tan\delta_{1}}{k^{3}}\boldsymbol{\nabla}\delta(\mathbf{r})\frac{\partial^{3}}{\partial r^{3}}r^{3}\mathbf{\boldsymbol{\nabla'},}\label{eq:pseu8}
\end{equation}
respectively. The symbol $\boldsymbol{\nabla}$ ($\boldsymbol{\nabla'}$)
acts only to the left (right) side of the pseudopotential.

\subsection{Effective range parameters}

For low energy scattering off a short range potential, it is possible
to express the variation of the phase shift as 
\begin{equation}
k^{2l+1}\cot\delta_{l}=-\frac{1}{a_{l}^{2l+1}}+\frac{1}{2}r_{l}k^{2}+O\left(k^{4}\right)\label{eq:pseu11}
\end{equation}
with only two parameters, the $l$-th partial wave scattering length
$a_{l}$ and the effective range $r_{l}$ \cite{Mott1965}. If we
define $R$ as the finite range of the potential, the effective range
$r_{l}$ can be expressed as \cite{Madsen2002}, 
\begin{equation}
r_{l}\propto-\frac{2\left(2l-1\right)!!}{\left(2l-1\right)R^{2l-1}}.\label{eq:pseu12}
\end{equation}
In the low energy limit we replace the term $\tan\delta_{l}/k^{2l+1}$in
Eqs. (\ref{eq:pseu7}) and (\ref{eq:pseu8}) by using Eq. (\ref{eq:pseu11}).
We then obtain the two-parameter pseudopotential for $\textit{s}$-wave
and $\textit{p}$-wave interactions as 
\begin{equation}
V_{s}=-\frac{2\pi\hbar^{2}}{\mu}\left(-a_{0}^{-1}+\frac{1}{2}r_{0}k^{2}\right)^{-1}\delta(\mathbf{r})\frac{\partial}{\partial r}r\label{eq:pseu15}
\end{equation}
and 
\begin{equation}
V_{p}=-\frac{\pi\hbar^{2}}{\mu}\left(-a_{1}^{-3}+\frac{1}{2}r_{1}k^{2}\right)^{-1}\boldsymbol{\nabla}\delta(\mathbf{r})\frac{\partial^{3}}{\partial r^{3}}r^{3}\boldsymbol{\nabla',}\label{eq:pseu16}
\end{equation}
respectively. 

Neglecting the term $r_{0}k^{2}/2$ in Eq. (\ref{eq:pseu15}), the
two-parameter s-wave pseudopotential will result in the famous Huang-Yang
$s$-wave zero-range pseudopotential,
\begin{equation}
\frac{2\pi\hbar^{2}a_{0}}{\mu}\delta(\mathbf{r})\frac{\partial}{\partial r}r.\label{eq:pseu13}
\end{equation}
However, as we know that using the single parameter, $s$-wave scattering
length, pseudopotential is not always a good approximation in a high-density
or tightly trapped system, as shown by by Blume and Bolda \textit{et
al}. \cite{Blume2002,Bolda2002}. In such cases, the next term in
the expansion needs to be included.

For higher partial-wave interactions, for example $p$-wave scattering,
the zero-range pseudopotential fails to describe the scattering even
in the low-energy limit \cite{Idziaszek}. This can be understood
by the zero-range approximation constraint condition $\left|1/a_{l}^{2l+1}\right|\gg r_{l}k^{2}$
or, using the result of the effective range equation (\ref{eq:pseu12}),
\begin{equation}
\left|ka_{l}\right|\ll\left(kR\right)^{\frac{2l-1}{2l+1}}.\label{eq:pseu14}
\end{equation}
Writing this out explicitly in the two cases of s-wave and p-wave
interactions, we see that this implies:
\begin{eqnarray}
\left|ka_{0}\right| & \ll & 1/\left(kR\right)\,\,\,\,\,\, s-wave\, case\nonumber \\
\left|ka_{1}\right| & \ll & \left(kR\right)^{1/3}\,\,\,\,\,\, p-wave\, case\label{eq:two-cases}
\end{eqnarray}
This means that zero-range approximation is useful for short range
interactions with $kR\rightarrow0$ in the s-wave case, since the
dimensionless interaction range $\left(kR\right)$ can always be made
arbitrarily small in order to satisfy the above inequality, even if
$\left|ka_{l}\right|\gg1$. In practise, this is achieved at low density
and ultralow temperatures, ie, by reducing $k$. However, this limit
cannot be used in the strongly interacting p-wave regime with $\left|ka_{l}\right|\gg1$.
For these higher partial-wave terms, the right-hand side will decrease
as $kR\rightarrow0$, meaning that a finite range correction is required
in order to reach strongly interacting regime. 

It should be noted that for any given values of range and scattering
length, one can always reach the zero-range regime at sufficient dilution,
as expected. In the s-wave case one has to satisfy $k^{2}\ll(R\left|a_{0}\right|)^{-1}$
. It is clearly possible to reach a regime where simultaneously $kR\ll1$
and $\left|ka_{0}\right|\gg1$, provided $k$ is small and $\left|a_{0}\right|$
is very large. However, in the p-wave case the inequality becomes
\[
k^{2}\ll R/\left|a_{1}\right|^{3}\,.
\]
The problem is that if $R\ll\left|a_{1}\right|$, which is the case
near a Feshbach resonance, then satisfying this p-wave inequality
requires low enough densities such that $\left|ka_{1}\right|\ll\left|kR\right|^{1/3}$
. At such low densities one is no longer in the strongly interacting
regime, since this inequality is only satisfied if $\left|ka_{1}\right|\ll1$.
Finally, it should be pointed out that the pseudopotential is not
a Hermitian operator. However, away from the unphysical region of
the origin, we have a well-defined scattering problem.

\section{two fermions in a harmonic trap}

The two-body problem with $s$-wave and $p$-wave interactions\cite{Suzuki2009}
in a three-dimensional harmonic trap was solved using a zero-range
pseudopotential by Busch and Idziaszek \textit{et al} \cite{Busch1998,Idziaszek2006}.
Here we use the two-parameter pseudopotential given above. In an axially
symmetric harmonic trap, the motion of the center of mass can be separated
from the relative motion, and the relative Hamiltonian is given by
\begin{equation}
\left[-\frac{\hbar^{2}}{2\mu}\nabla^{2}+\frac{1}{2}\mu\omega^{2}\left(\eta^{2}\boldsymbol{\rho}^{2}+\mathbf{z}^{2}\right)+V_{ps}\right]\psi_{rel}=E_{rel}\psi_{rel},\label{eq:twobody1}
\end{equation}
where $\boldsymbol{\rho}=\boldsymbol{\rho}_{1}-\boldsymbol{\rho}_{2}$,
$\mathbf{z}=\mathbf{z}_{1}-\mathbf{z}_{2}$ and $\mathbf{r}=\mathbf{r}_{1}-\mathbf{r}_{2}$
are the relative coordinates, $\mu$ is the reduced mass, $\omega$
is the transverse frequency of the trap, $\eta=\omega_{\rho}/\omega$,
and $V_{ps}(\mathbf{r})$ is the two-parameter pseudopotential. To
solve Eq. (\ref{eq:twobody1}), we expand the relative wavefunction
$\psi_{rel}\left(\mathbf{r}\right)$ into the complete set of the
eigenfunctions of three-dimensional harmonic oscillator,
\begin{equation}
\psi_{rel}\left(\mathbf{r}\right)=\sum_{n_{1},n_{2},n_{3}}C_{n_{1}n_{2}n_{3}}\varphi_{n_{1,}n_{2}n_{3}}\mathbf{\left(r\right),}\label{eq:twobody2}
\end{equation}
where $\varphi_{n_{1}n_{2}n_{3}}(\mathbf{r})=\phi_{n_{1}}(\eta x/d)\phi_{n_{2}}(\eta y/d)\phi_{n_{3}}(z/d)$,
$d=\sqrt{\hbar/\mu\omega}$ is the oscillator length in the transverse
direction and $\phi_{n}(\xi)$ can be expressed by a Hermite polynomial
as
\begin{gather*}
\phi_{n}(\xi)=\sqrt{\frac{1}{\sqrt{\pi}2^{n}n!}}e^{-\xi^{2}/2}H_{n}\left(\xi\right).
\end{gather*}
Considering an isotropic harmonic trap, $\varphi_{n_{1}n_{2}n_{3}}(\mathbf{r})$
can be replaced by the spherical harmonic function $Y_{lm}\left(\theta,\varphi\right)$
and the function $R_{nl}\left(r\right)$, written as 
\begin{equation}
\varphi_{nlm}\left(\mathbf{r}\right)=R_{nl}\left(r\right)Y_{lm}\left(\theta,\varphi\right).\label{eq:twobody3}
\end{equation}
Using the generalized Laguerre polynomials $L_{n}^{(l+1/2)}$, we
can express the function $R_{nl}$ as $R_{nl}=N_{nl}\left(r/d\right)^{l}exp\left(-r^{2}/2d^{2}\right)L_{n}^{(l+1/2)}\left(r^{2}/d^{2}\right)$.
Here $N_{nl}$ is the normalization coefficient.

\subsection{$s$-wave interaction in a 3D isotropic trap}

As a preliminary calculation, we consider two fermions with unlike
spins interacting via $s$-wave interactions in a 3D isotropic harmonic
trap, with $\eta=\omega_{z}/\omega=1$. The relative wavefunction
can be written in the form of Eq. (\ref{eq:twobody3}). For $s$-wave
interactions, we only need to keep $l=0$ in Eq. (\ref{eq:twobody3})
as $l\neq0$ modes are not affected by the interactions. The relative
wavefunction of interest can then be written as 
\begin{equation}
\psi_{rel}\left(\mathbf{r}\right)=\sum_{n}C_{n}\varphi_{n}\mathbf{\left(r\right)},\label{eq:swave}
\end{equation}
where $\varphi_{n}\left(\mathbf{r}\right)=R_{n0}\left(r\right)Y_{00}\left(\theta,\varphi\right)$.
We insert this relative wavefunction into Eq. (\ref{eq:twobody1}).
Following the derivation by Busch \textit{et al.} \cite{Busch1998}
the energy level $E_{rel}$ should satisfy the following expression,

\begin{gather}
\frac{2\pi\hbar\omega}{d}\left[\frac{\partial}{\partial r}\left(r\sum_{n}\frac{\varphi_{n}^{*}\left(0\right)\varphi_{n}\left(\mathbf{r}\right)}{E_{n}-E_{rel}}\right)\right]_{r\rightarrow0}=-\frac{1}{a_{0}}+\frac{1}{2}r_{0}k^{2},\label{eq:swaveeigenvalue}
\end{gather}
where $E_{n}=\left(2n+3/2\right)\hbar\omega$ and we have defined
$E_{rel}=\left(2\nu+3/2\right)\hbar\omega$. By using $R_{n0}=N_{n0}exp\left(-r^{2}/2d^{2}\right)L_{n}^{1/2}\left(r^{2}/d^{2}\right)$
and the relationship between the Laguerre polynomials $L_{n}^{1/2}\left(x\right)$
and the confluent hypergeometric function $U,$ 
\begin{gather}
\sum_{n=0}^{\infty}\frac{L_{n}^{1/2}\left(x\right)}{n-\nu}=\Gamma\left(-\nu\right)U\left(-\nu,\frac{3}{2},x\right),\label{eq:LP}
\end{gather}
Eq. (\ref{eq:swaveeigenvalue}) can be simplified to
\begin{gather}
\frac{1}{\sqrt{\pi}}\Gamma\left(-\nu\right)U\left(-\nu,\frac{3}{2},\frac{r^{2}}{d^{2}}\right)=-\frac{d}{a_{0}}+\frac{d}{2}r_{0}k^{2},\label{eq:swaveeg}
\end{gather}
where $\Gamma\left(x\right)$ is the Gamma function. By examining
the short-range behavior of the confluent hypergeometric function,
\begin{equation}
\Gamma\left(-\nu\right)U\left(-\nu,\frac{3}{2},r^{2}\right)=-\sqrt{\pi}\left[\frac{2\Gamma\left(-\nu\right)}{\Gamma\left(-\nu-1/2\right)}-\frac{1}{r}+O\left(r\right)\right],\label{eq:twobody7}
\end{equation}
and taking the approximation $k^{2}=2\mu E_{rel}/\hbar^{2}$, we obtain
a secular equation for the relative energy levels: 
\begin{equation}
\frac{2\Gamma\left(-\nu\right)}{\Gamma\left(-\nu-1/2\right)}=\frac{d}{a_{0}}-\frac{r_{0}}{d}\cdot\left(2\nu+\frac{3}{2}\right).\label{eq:twobody5}
\end{equation}
Here the second term on the right hand side of Eq. (\ref{eq:twobody5})
shows the effects of the finite-range interaction potential. Taking
a zero range limit, with $r_{0}=0$, Eq. (\ref{eq:twobody5}) gives
Busch's earlier result \cite{Busch1998}. The relative wave functions
can be found to be: 
\begin{equation}
\psi_{rel}\left(\mathbf{r}\right)\propto exp\left(-\frac{r^{2}}{2d^{2}}\right)\Gamma\left(-\nu\right)U\left(-\nu,\frac{3}{2},\frac{r^{2}}{d^{2}}\right).\label{eq:twobody6}
\end{equation}

In Fig. 1, we give the energy spectrum of two fermions with $s$-wave
interactions as a function of the dimensionless interaction strength
$d/a_{0}$. Here we consider two cases: 
\begin{description}
\item [{(i)}] zero-range limit (red dotted line) and 
\item [{(ii)}] $r_{0}/d=2R/d=0.1$ (blue solid line). 
\end{description}
It is easy to see that, except for the lowest bound state, the two
lines differ slightly. The most significant difference appears around
the unitarity limit with $a_{0}\rightarrow\pm\infty$. Thus, the zero-range
approximation appear to be sound as far as the lowest bound state
is concerned. Therefore, the scattering length $a_{0}$ can be used
to characterize the low-energy$s$-wave scattering, provided that
the finite range of the interaction potential is smaller than the
characteristic length of the harmonic trap, $d$.

In the limiting case of zero scattering length, we find that the asymptotic
behavior of the energy spectrum for the $n$-th level can be described
by,
\begin{gather*}
\frac{E_{rel}}{\hbar\omega}=2n+\frac{3}{2}+\frac{4\Gamma\left(n+3/2\right)}{\pi\Gamma\left(n+1\right)\left[d/a_{0}-\left(2n+3/2\right)r_{0}/d\right]},
\end{gather*}
where $n=0,1,2,\ldots$ is a non-negative integer. 

Near the unitarity limit, the asymptotic behavior of the energy spectrum
for the $n$-th level can be written as 
\begin{equation}
\frac{E_{rel}}{\hbar\omega}=2n+\frac{1}{2}+2q,\label{eq:twobody8}
\end{equation}
where
\begin{equation}
q=-\frac{\Gamma\left(n+1/2\right)d}{2\pi\Gamma\left(n+1\right)a_{0}}+\frac{\left(2n+1/2\right)\Gamma\left(n+1/2\right)r_{0}}{2\pi\Gamma\left(n+1\right)d}.\label{eq:1.2}
\end{equation}

\begin{figure}
\includegraphics[width=0.5\textwidth]{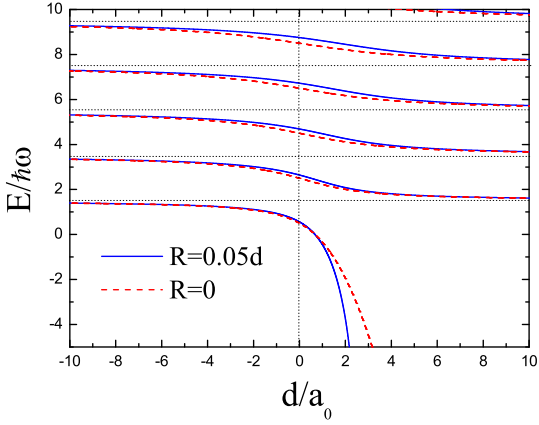}

\caption{(Color online) The energy spectrum of two fermions with $s$-wave
interactions in a 3D isotropic harmonic trap, as a function of the
dimensionless interaction parameter $d/a_{0}$ at two different effective
ranges of interactions: $r_{0}/d=0$ (red dotted lines) and $r_{0}/d=0.1$
(blue solid lines).}

\label{Flo:fig1}
\end{figure}

\subsection{$p$-wave interactions in a 3D isotropic trap }

Let us now consider two fermions with $p$-wave interactions in a
3D isotropic harmonic trap. We keep only the $l=1$ term in Eq. (\ref{eq:twobody3}),
since other $l\neq1$ channels are not affected by $p$-wave interactions.
We may then write the relative wavefunction, 
\begin{gather}
\psi_{rel}\left(\mathbf{r}\right)=\sum_{nm}C_{n1m}R_{n1}\left(r\right)Y_{1m}\left(\theta,\varphi\right).\label{eq:pwave}
\end{gather}
Substituting Eq. (\ref{eq:pwave}) into Eq. (\ref{eq:twobody1}),
we obtain
\begin{gather}
\sum_{nm}\frac{\left\{ \boldsymbol{\nabla}\varphi_{n1m}^{*}\left(\mathbf{r}\right)\left(\partial^{3}/\partial r^{3}\right)\left[r^{3}\boldsymbol{\nabla}\varphi_{n1m}\left(\mathbf{r}\right)\right]\right\} {}_{r\rightarrow0}}{E_{nl}-E_{rel}}\nonumber \\
=\frac{\mu}{\pi\hbar^{2}}\left(-a_{1}^{-3}+\frac{1}{2}r_{1}k^{2}\right)^{-1},\label{eq:6.1}
\end{gather}
 where $E_{nl}=\left(2n+1+3/2\right)\hbar\omega$ is the non-interacting
relative energy and we have written the interacting energy eigenvalue
as $E_{rel}=\left(2\nu+5/2\right)\hbar\omega$. Using $\sum_{m}Y_{lm}^{*}\left(\theta,\varphi\right)Y_{lm}\left(\theta,\varphi\right)=\left(2l+1\right)/4\pi$
and $R_{n1}=N_{n1}\left(r/d\right)exp\left(-r^{2}/2d^{2}\right)L_{n}^{3/2}\left(r^{2}/d^{2}\right)$,
we obtain
\begin{gather}
\left\{ \frac{\partial^{3}}{\partial r^{3}}r^{2}\left[r\exp\left(-\frac{r^{2}}{2d^{2}}\right)\sum_{n}\frac{L_{n}^{3/2}\left(r^{2}/d^{2}\right)}{\left(2n+5/2\right)\hbar\omega-E_{rel}}\right]\right\} _{r\rightarrow0}\nonumber \\
=\frac{d^{5}\sqrt{\pi}\mu}{2\hbar^{2}}\left(-a_{1}^{-3}+\frac{1}{2}r_{1}k^{2}\right)^{-1}.\label{eq:6.2}
\end{gather}
Using the identity between the Laguerre polynomials $L_{n}^{3/2}\left(r^{2}\right)$
and the confluent hypergeometric function $U$,
\begin{equation}
\sum_{n}\frac{L_{n}^{3/2}\left(r^{2}/d^{2}\right)}{\left(2n+5/2\right)\hbar\omega-E_{rel}}=\frac{1}{2\hbar\omega}\Gamma\left(-\nu\right)U\left(-\nu,\frac{5}{2},\frac{r^{2}}{d^{2}}\right),\label{eq:6.3}
\end{equation}
and the asymptotic behavior of the confluent hypergeometeric function
at $r\rightarrow0$,
\begin{multline}
\frac{1}{\pi}\Gamma\left(-\nu\right)U\left(-\nu,\frac{5}{2},r^{2}\right)=-\frac{r^{-3}}{\Gamma\left(-1/2\right)}\\
-\frac{2\nu+3}{r\Gamma\left(-1/2\right)}+\frac{\Gamma\left(-\nu\right)}{\Gamma\left(-\nu-3/2\right)\Gamma\left(5/2\right)},\label{eq:6.4}
\end{multline}
the eigenvalue equation Eq. (\ref{eq:6.2}) leads to
\begin{equation}
\frac{\Gamma\left(-\nu+1/2\right)}{\Gamma\left(-\nu-1\right)}=-\frac{1}{8}\cdot\left(\frac{d}{a_{1}}\right)^{3}+\frac{1}{8}\cdot\left(dr_{1}\right)\cdot\left(2\nu+\frac{3}{2}\right).\label{eq:twobody9}
\end{equation}
Here we take the same approximation $k^{2}=2\mu E_{rel}/\hbar^{2}$
as in the $s$-wave case. The corresponding eigenfunctions have the
form, 
\begin{eqnarray}
\psi_{rel}\left(\mathbf{r}\right) & \propto & \left(\frac{r}{d}\right)exp\left(-\frac{r^{2}}{2d^{2}}\right)\times\nonumber \\
\begin{array}{c}
\end{array} &  & \Gamma\left(-\nu+\frac{1}{2}\right)U\left(-\nu+\frac{1}{2},\frac{5}{2},\frac{r^{2}}{d^{2}}\right)Y_{1m}.\label{eq:6.5}
\end{eqnarray}

In Fig. 2, we give the energy spectrum of two fermions with $p$-wave
interactions as a function of the dimensionless inverse scattering
volume $d^{3}/a_{1}^{3}$. The red dotted line and blue solid line
correspond to a zero effective range $r_{1}d=0$ and a finite effective
range $r_{1}d=-40$, respectively. The value of $r_{1}d=-40$ is calculated
from the finite range of interaction potentials, $R=0.05d$, using
$R=-2/r_{1}$. Unlike the $s$-wave case, near the unitarity limit,
a small finite-range parameter in the interaction potential can induce
a significant change of the energy spectrum, implying that the use
of zero-range pseudopotential is not justified for $p$-wave interactions
in the unitarity limit, as expected. 

This can be understood using Eq. (\ref{eq:pseu14}). For $p$-wave
interactions, the zero-range approximation is valid only if $\left|ka_{1}\right|\ll\left(kR\right)^{1/3}$.
For a small value of $R$ and a large scattering length $a_{1}$,
the wavelength or the collision energy therefore should be extremely
small in order to satisfy this constraint. As explained in Section
II, this limit can only be achieved if $\left|ka_{1}\right|\ll1$
, which is no longer in the strongly-interacting regime. In contrast,
for $s$-wave interactions, we only require that $\left|ka_{0}\right|\ll\left(kR\right)^{-1}$,
which can generally be satisfied at low dilutions near a Feshbach
resonance. We have checked our two-parameter pseudopotential method,
by comparing its energy spectrum with that predicted by a more complex
energy-dependent pseudopotential approach \cite{Idziaszek}. For typical
parameters corresponding to the experimental condition for $^{40}$K
fermions, we find an excellent agreement between these two pseudopotentials.

In the non-interacting limit of zero scattering length, we find that
the following asymptotic form for the $n$-th energy level, 
\begin{gather}
\frac{E_{rel}}{\hbar\omega}=2n+\frac{5}{2}+\frac{8\Gamma\left(n+5/2\right)}{\pi\Gamma\left(n+1\right)\left(d^{3}/2a_{1}^{3}-r_{1}d\left(n+5/4\right)\right)}.\label{eq:6.6}
\end{gather}
In the zero range limit, Eq. (\ref{eq:6.6}) implies that the interacting
energy spectrum differs mostly from the non-interacting spectrum at
the resonance position $a_{1}\rightarrow\infty$. However, the existence
of a finite range of interactions will shift the position to the BCS
side, which is determined by the condition, $d^{3}/(2a_{1}^{3})-r_{1}d\left(n+5/4\right)=0$.

Near the unitarity limit, the lowest energy level is approximately
zero. The asymptotic value of energy spectrum for the $n$-th level
is given by, 
\begin{equation}
\frac{E_{rel}}{\hbar\omega}=2n+\frac{5}{2}+2q,\label{eq:1.3}
\end{equation}
where
\begin{equation}
q=\frac{-\Gamma\left(n+\frac{5}{2}\right)}{\pi\left(n+\frac{5}{4}\right)\Gamma\left(n+1\right)}\left(\frac{4}{r_{1}d}+\frac{2d}{a_{1}^{3}r_{1}^{2}\left(n+5/4\right)}\right).\label{eq:1.4}
\end{equation}

\begin{figure}
\includegraphics[width=0.5\textwidth]{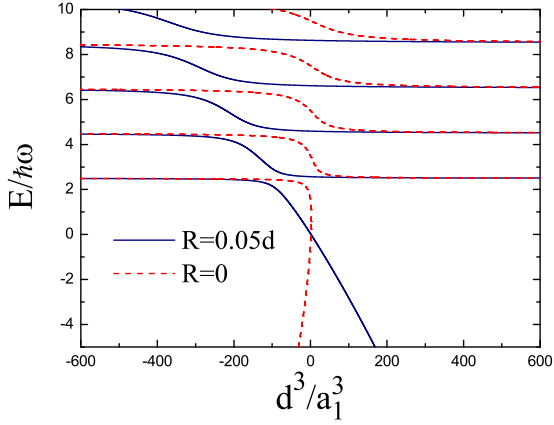}

\caption{(Color online) The energy spectrum of two fermions with $p$-wave
interactions in a 3D isotropic harmonic trap, as a function of the
dimensionless inverse scattering volume $d^{3}/a_{1}^{3}$ at two
different effective ranges: $r_{1}d=0$ (red dotted lines) and $r_{1}d=-40$
(blue solid lines).}

\label{Flo:fig2}
\end{figure}

\subsection{$p$-wave interactions and 3D anisotropic traps}

Let us now turn to the case of two fermions with $p$-wave interactions
in an \emph{anisotropic} 3D harmonic trap $\mu\omega^{2}\left(\eta^{2}\boldsymbol{\rho}^{2}+\mathbf{z}^{2}\right)/2$.
We have to take the general form Eq. (\ref{eq:twobody2}) for the
relative wavefunction $\psi_{rel}\left(\mathbf{r}\right)$. By substituting
Eq. (\ref{eq:twobody2}) into Eq.(\ref{eq:twobody1}), we follow the
derivation for $s$-wave interactions given by Busch \textit{et al}.
\cite{Busch1998}. This leads to the following expression for the
eigenvalues $E_{rel}$,
\begin{gather}
\sum_{n_{1}^{\prime}n_{2}^{\prime}n_{3}^{\prime}}C_{n_{1}^{\prime}n_{2}^{\prime}n_{3}^{\prime}}\frac{\left\{ \boldsymbol{\nabla}\varphi_{n'_{1}n'_{2}n'_{3}}^{*}\left(\partial^{3}/\partial r^{3}\right)\left[r^{3}\boldsymbol{\nabla}\varphi_{n'_{1}n'_{2}n'_{3}}\mathbf{\left(r\right)}\right]\right\} {}_{r\rightarrow0}}{\left(E_{n'_{1}n'_{2}n'_{3}}-E_{rel}\right)C_{n_{1}n_{2}n_{3}}}\nonumber \\
=\frac{\mu}{\pi\hbar^{2}}\left(-a_{1}^{-3}+\frac{1}{2}r_{1}k^{2}\right),\label{eq:anpwev}
\end{gather}
where $C_{n_{1}n_{2}n_{3}}=\left[\nabla\varphi_{n'_{1,}n'_{2}n'_{3}}^{*}\mathbf{\left(r\right)}\right]_{r\rightarrow0}$
and the non-interacting energy spectrum is given by $E_{n_{1}n_{2}n_{3}}=\left(n_{1}+1/2\right)\hbar\omega_{x}+\left(n_{2}+1/2\right)\hbar\omega_{y}+\left(n_{3}+1/2\right)\hbar\omega_{z}$.
To proceed, we re-write the summation in the above equation by using
the identity, 
\begin{gather*}
\frac{1}{E_{n}-E}=\int_{0}^{\infty}dte^{-t\left(E_{n}-E\right)}
\end{gather*}
for $E<E_{0}=\hbar\left(\omega_{x}+\omega_{y}+\omega_{z}\right)/2$,
and use the relation satisfied by the Hermite polynomial, 
\begin{gather*}
\sum_{k=0}^{\infty}\frac{t^{k}}{2^{k}k!}H_{k}(x)H_{k}(y)=\frac{e^{\left(2txy-t^{2}x^{2}-t^{2}y^{2}\right)/\left(1-t^{2}\right)}}{\sqrt{1-t^{2}}}.
\end{gather*}
As a result, the left-hand side of Eq. (\ref{eq:anpwev}) can be expressed
by using the function ($l=x,y,z$ )
\[
H_{l}\left(\epsilon,\mathbf{r}\right)=\int_{0}^{\infty}dt\frac{e^{t\left(\epsilon-\eta_{l}\right)}}{1-e^{-2\eta_{l}t}}\prod_{i=1}^{3}\frac{\exp\left[-\eta_{i}\frac{x_{i}^{2}e^{-2\eta_{i}t}}{1-e^{-2\eta_{i}t}}\right]}{\sqrt{1-e^{-2\eta_{i}t}}}.
\]
At $r\rightarrow0$, the asymptotic behavior of $H_{l}\left(\epsilon=\left(E-E_{0}\right)/\hbar\omega,\mathbf{r}\right)$
is given by,
\[
H_{l}\left(\epsilon,\mathbf{r}\right)=F_{l}\left(\epsilon,\mathbf{r}\right)+\frac{\sqrt{\pi}}{4\eta\eta_{l}r^{3}}+\frac{\left(2\epsilon+2\eta+1\right)\sqrt{\pi}}{8\eta\eta_{l}r},
\]
where
\begin{eqnarray}
F_{l}\left(\epsilon,\mathbf{r}\right) & = & \int_{0}^{\infty}dt\left\{ \frac{e^{t\left(\epsilon-\eta_{l}\right)}}{1-e^{-2\eta_{l}t}}\prod_{i=1}^{3}\frac{\exp\left[-\eta_{i}\frac{x_{i}^{2}e^{-2\eta_{i}t}}{1-e^{-2\eta_{i}t}}\right]}{\sqrt{1-e^{-2\eta_{i}t}}}-\right.\nonumber \\
 &  & \left.-\left[\frac{e^{-r^{2}/2t}}{4\sqrt{2}\eta\eta_{l}}\left(\frac{1}{t^{5/2}}+\frac{2\epsilon+2\eta+1}{2t^{3/2}}\right)\right]\right\} \label{eq:Fl}
\end{eqnarray}
and $\eta_{x}=\eta_{y}=\eta$ and $\eta_{z}=1$. After a straightforward
but lengthy calculation, we find that

\begin{equation}
\frac{12\eta\eta_{l}}{\sqrt{\pi}}F_{l}\left(\frac{E_{rel}-E_{0}}{\hbar\omega},0\right)=-\left(\frac{d}{a_{1}}\right)^{3}+r_{1}d\frac{E_{rel}}{\hbar\omega}.\label{eq:twobody13}
\end{equation}
Here we take the same approximation $k^{2}=2\mu E_{rel}/\hbar^{2}$
as before and $F_{l}\left(E_{rel},0\right)\equiv F_{l}\left(E_{rel},\mathbf{r}=0\right)$.
In an axially symmetric trap, we may use the good quantum number of
projected angular momentum $m$ to label the energy levels. The secular
equations for $m=\pm1$ and for $m=0$ are, respectively \cite{Idziaszek},
\begin{equation}
\frac{12\eta^{2}}{\sqrt{\pi}}F_{x}\left(\frac{E_{rel}-E_{0}}{\hbar\omega},0\right)=-\frac{d^{3}}{a_{1}^{3}}+r_{1}d\cdot\frac{E_{rel}}{\hbar\omega}\label{eq:twobody15}
\end{equation}
and 

\begin{equation}
\frac{12\eta}{\sqrt{\pi}}F_{z}\left(\frac{E_{rel}-E_{0}}{\hbar\omega},0\right)=-\frac{d^{3}}{a_{1}^{3}}+r_{1}d\cdot\frac{E_{rel}}{\hbar\omega}.\label{eq:twobody16}
\end{equation}
To calculate the eigenvalue, for $E_{rel}<E_{0}$ we use the integral
equation Eq. (\ref{eq:Fl}), while for $E_{rel}>E_{0}$ we use the
recurrence relations \cite{Idziaszek},
\begin{flalign}
 & F_{x}\left(\epsilon+2\eta\right)-2F_{x}\left(\epsilon\right)+F_{x}\left(\epsilon-2\eta\right)\nonumber \\
 & =\frac{\sqrt{\pi}}{2}\cdot\frac{\Gamma\left(-\eta/2-\epsilon/2\right)}{\Gamma\left(1/2-\eta/2-\epsilon/2\right)}\label{eq:twobody19}
\end{flalign}
and 
\begin{equation}
F_{z}\left(\epsilon+2\eta,0\right)-F_{z}\left(\epsilon,0\right)=-\sqrt{\pi}\cdot\frac{\Gamma\left(1/2-\eta-\epsilon/2\right)}{\Gamma\left(-\eta-\epsilon/2\right)}\label{eq:twobody20}
\end{equation}
where $\epsilon=\left(E-E_{0}\right)/\hbar\omega$. 

In Figs. 3 and 4, we show the energy spectrum of two interacting fermions
with $p$-wave interactions, as a function of the dimensionless inverse
scattering volume $d^{3}/a_{1}^{3}$ at an effective range $r_{1}d=-40$,
in a cigar-shaped trap ($\eta=5$) and in a pancake-shaped trap ($\eta=0.2$),
respectively. These plots are similar to the energy spectrum of two
fermions in an isotropic trap. However, due to the trap anisotropy,
the degeneracy for different projected angular momentum $m$ is removed.

\begin{figure}
\includegraphics[clip,width=0.5\textwidth]{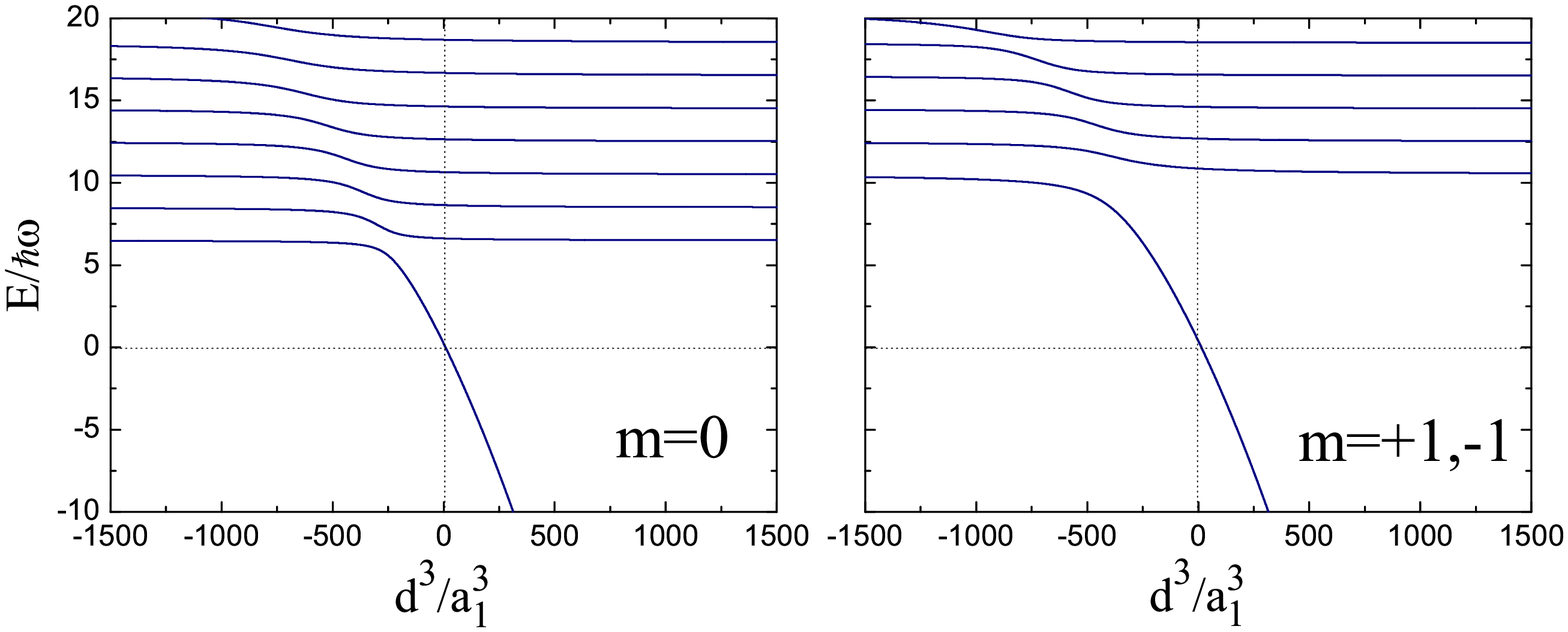}

\caption{(Color online) The energy spectrum of two fermions with $p$-wave
interactions in a 3D anisotropic harmonic trap, as a function of the
dimensionless inverse scattering volume $d^{3}/a_{1}^{3}$ at $\eta=\omega_{\rho}/\omega=5$
and the effective range $r_{1}d=-40$. }

\label{Flo:fig3}
\end{figure}

\begin{figure}
\includegraphics[width=0.5\textwidth]{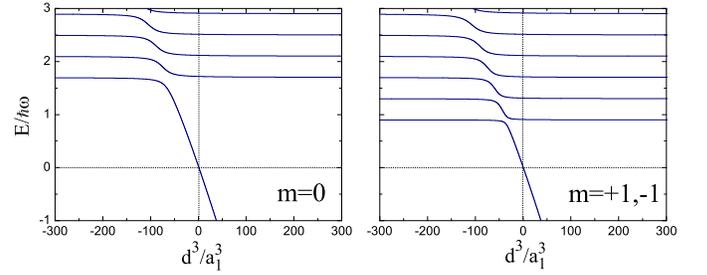}

\caption{(Color online) The energy spectrum of two fermions with $p$-wave
interactions, as in Fig. 3, but with $\eta=\omega_{\rho}/\omega=0.2$.}

\label{Flo:fig4}
\end{figure}

\section{Virial Expansion of strongly correlated fermions}

The knowledge of few-particle exact solutions provides a useful input
for investigating the high-temperature behavior and many-body physics
of a strongly correlated quantum gas. This is provided by the quantum
virial expansion technique \cite{HO,Liu2009,HU2010,Liu2010}. The
essential idea of the quantum virial expansion is that at high temperatures
the chemical potential $\mu$ is negative and the fugacity $z\equiv\exp(\mu/k_{B}T)\equiv\exp(\beta\mu)\ll1$
is a well-defined small parameter. We can therefore expand the thermodynamic
potential $\Omega$ of a quantum system in powers of the fugacity,
regardless of the strength of the interactions. In general we may
write,

\begin{equation}
\Omega=-k_{B}TQ_{1}\left(z+b_{2}z^{2}+\cdots+b_{n}z^{n}+\cdots\right)\label{eq:virial1}
\end{equation}
where $Q_{n}$ is the partition function of a cluster that contain
$n$ particles 
\begin{equation}
Q_{n}=\mathrm{Tr}_{n}\left[exp\left(-\mathcal{H}/k_{B}T\right)\right].\label{eq:virial2}
\end{equation}
The trace $Tr_{n}$ takes into account all the $n$-particle states
with a proper symmetry. The $n$-th virial coefficient $b_{n}$ has
the form:
\begin{eqnarray}
b_{2} & = & \frac{Q_{2}-Q_{1}^{2}/2}{Q_{1}}\label{eq:virial6}\\
b_{3} & = & \frac{Q_{3}-Q_{1}Q_{2}+Q_{1}^{3}/3}{Q_{1}}\label{eq:virial7}\\
 & \vdots\nonumber 
\end{eqnarray}
All the other thermodynamic properties can then be derived from $\Omega$
via the standard thermodynamic relations.

It is convenient to focus on the effect of interactions on the virial
coefficients. To this end, we consider the differences $\Delta Q_{n}=Q_{n}-Q_{n}^{(1)}$
and $\Delta b_{n}=b_{n}-b_{n}^{(1)}$, where the superscript {}``$1$''
denotes an ideal, non-interacting system having the same fugacity.
Accordingly, we rewrite the thermodynamic potential in the form, 
\begin{equation}
\Omega=\Omega^{(1)}-k_{B}TQ_{1}\left[\Delta b_{2}z^{2}+\cdots+\Delta b_{n}z^{n}+\cdots\right],\label{omega}
\end{equation}
 where $\Omega^{(1)}$ is the non-interacting thermodynamic potential
and
\begin{eqnarray}
\Delta b_{2} & = & \Delta Q_{2}/Q_{1}\label{eq:virial8}\\
\Delta b_{3} & = & \Delta Q_{3}/Q_{1}-\Delta Q_{2}\label{eq:virial9}\\
 & \vdots\nonumber 
\end{eqnarray}
We now describe how to calculate the non-interacting thermodynamic
potential $\Omega^{(1)}$ and the virial coefficients $\Delta b_{n}$.

\subsection{Ideal thermodynamic potential}

The non-interacting virial coefficients can be determined straightforwardly
from the non-interacting thermodynamic potential. For a harmonically
trapped Fermi gas, the non-interacting thermodynamic potential in
the semiclassical limit (\textit{i.e.}, neglecting the discreteness
of the spectrum) is, 
\begin{equation}
\Omega^{(1)}=-\frac{2\left(k_{B}T\right)^{4}}{\left(\hbar\bar{\omega}\right)^{3}}\frac{1}{2}\int\limits _{0}^{\infty}t^{2}\ln\left(1+ze^{-t}\right)dt,\label{non_interacting_Omega_Trap}
\end{equation}
where $Q_{1}=2\left(k_{B}T\right)^{3}/\left(\hbar\bar{\omega}\right)^{3}$
and $\hbar\bar{\omega}=\eta^{2/3}\hbar\omega$. Taylor-expanding the
non-interacting thermodynamic potential in powers of $z$ gives rise
to 
\begin{equation}
b_{n}^{(1)}=\frac{\left(-1\right)^{n+1}}{n^{4}}.
\end{equation}

\subsection{Second virial coefficient in a harmonic trap }

To calculate the second virial coefficient of a trapped interacting
Fermi gas, we use $\tilde{\omega}=\hbar\omega/k_{B}T\ll1$ to characterize
the intrinsic length scale relative to the trap. To obtain $\Delta b_{2}$,
we consider separately $\Delta Q_{2}$ and $Q_{1}$. The single-particle
partition function $Q_{1}$ can be determined by the single-particle
spectrum of a 3D harmonic oscillator or alternatively can be derived
from the well-known ideal thermodynamic potential. We find 
\begin{equation}
Q_{1}\cong2\left(k_{B}T\right)^{3}/\left(\hbar\bar{\omega}\right)^{3}\,.
\end{equation}
The prefactor of two accounts for the two possible spin states of
a single fermion. In the calculation of $\Delta Q_{2}$, it is easy
to see that the summation over the center-of-mass energy level gives
exactly $Q_{1}/2$. Thus, the second virial coefficient is determined
entirely by the relative energy spectrum $E_{rel}$ and we find that,
\begin{equation}
\Delta b_{2}=\frac{1}{2}\sum\left[\exp(-E_{rel})-\exp(-E_{rel}^{\left(1\right)})\right],\label{db2}
\end{equation}
where $E_{rel}^{\left(1\right)}$ is the relative energy for non-interacting
two-fermions and the summation is over the whole energy spectrum.
In the following, we focus on the most interesting case of the unitary
limit.

\subsubsection{$s$-wave interaction }

Let us first consider the simplest case of $s$-wave interactions.
At resonance with an infinitely large scattering length and zero effective
range of interactions, the spectrum is known exactly: $E_{rel}=\left(2n+1/2\right)\hbar\omega$,
giving rise to, 
\begin{equation}
\Delta b_{2}=\frac{1}{2}\frac{\exp\left(-\tilde{\omega}/2\right)}{\left[1+\exp\left(-\tilde{\omega}\right)\right]}=+\frac{1}{4}-\frac{1}{32}\tilde{\omega}^{2}+\cdots.\label{db2attTrap}
\end{equation}
The leading term $1/4$ on the right-hand side of the above equation
is $universal$ and temperature independent. This is anticipated following
the universality argument first suggested by Ho \cite{HoUniversality}.
The term $\tilde{\omega}^{2}$ in Eqs. (\ref{db2attTrap}) is \emph{non-universal}
and is related to the intrinsic length scale of the harmonic trap.
However, it is negligibly small for a Fermi cloud with a large number
of atoms.

The existence of a finite range interaction term causes another non-universal
correction. Near resonance, the relative energy levels are given by
Eq. (\ref{eq:twobody8}), 
\begin{equation}
E_{rel}=\left(2n+\frac{1}{2}+2q\right)\hbar\omega\,,
\end{equation}
where 
\begin{equation}
q=-\frac{\Gamma\left(n+1/2\right)d}{2\pi\Gamma\left(n+1\right)a_{0}}+\frac{\left(2n+1/2\right)\Gamma\left(n+1/2\right)r_{0}}{2\pi\Gamma\left(n+1\right)d}.
\end{equation}
As $q\ll1$ for a small effective-range of interactions near unitarity,
we have 
\begin{gather}
\exp\left[-\tilde{\omega}\left(2n+\frac{1}{2}+2q\right)\right]\simeq\left(1+2q\tilde{\omega}\right)\exp\left[-\tilde{\omega}\left(2n+\frac{1}{2}\right)\right].\label{eq:5.1}
\end{gather}
Using 
\begin{gather}
\sum_{n=0}^{\infty}\frac{\Gamma\left(n+\frac{1}{2}\right)}{\Gamma\left(n+1\right)}\exp\left[-2n\tilde{\omega}\right]\thickapprox\sqrt{\frac{\pi}{2}}\cdot\frac{1}{\sqrt{\tilde{\omega}}}
\end{gather}
and
\begin{gather}
\sum_{n=0}^{\infty}\frac{\left(2n+\frac{1}{2}\right)\Gamma\left(n+\frac{1}{2}\right)}{\Gamma\left(n+1\right)}\exp\left[-2n\tilde{\omega}\right]\thickapprox\frac{\sqrt{\pi}}{2\sqrt{2}}\cdot\frac{1}{\tilde{\omega}^{3/2}},
\end{gather}
we obtain
\begin{gather}
\sum_{n=0}^{\infty}\left(\frac{\hbar\omega}{k_{B}T}\cdot2q\right)\exp\left[-\frac{\hbar\omega}{k_{B}T}\left(2n+\frac{1}{2}\right)\right]\nonumber \\
\approx-\frac{1}{\sqrt{2\pi}}\cdot\sqrt{\tilde{\omega}}\cdot\frac{d}{a_{0}}+\frac{1}{2\sqrt{2\pi}}\cdot\frac{1}{\sqrt{\tilde{\omega}}}\cdot\frac{r_{0}}{d}.\label{eq:5.9}
\end{gather}
Thus, the second virial coefficient near resonance can be analytically
written as
\begin{gather}
\Delta b_{2,s}=\frac{1}{4}-\frac{\tilde{\omega}^{2}}{32}+\frac{1}{\sqrt{2\pi\tilde{T}}}\frac{1}{k_{F}a_{0}}-\frac{k_{F}r_{0}\sqrt{\tilde{T}}}{8\sqrt{2\pi}},\label{eq:5.10}
\end{gather}
where $\tilde{T}=T/T_{F}$ is the reduced temperature, $k_{F}=\sqrt{2m\omega/\hbar}\left(3N\right)^{1/6}$
and $T_{F}=\left(3N\right)^{1/3}\hbar\omega/k_{B}$ are Fermi momentum
and Fermi temperature, respectively. The subscript {}``s'' stands
for the $s$-wave interaction.

\subsubsection{$p$-wave interaction}

For 3D \emph{isotropic} harmonic traps, the relative energy spectrum
for near resonant $p$-wave interactions is described by Eq. (\ref{eq:1.3}),
\[
E_{rel}=\left(2n+\frac{5}{2}+2q\right)\hbar\omega,
\]
where 
\[
q=\frac{-\Gamma\left(n+5/2\right)}{\pi\left(n+5/4\right)\Gamma\left(n+1\right)}\left(\frac{4}{r_{1}d}+\frac{2d}{a_{1}^{3}r_{1}^{2}\left(n+5/4\right)}\right).
\]
Note that the lowest energy level is approximately zero. In the limits
of $\left(r_{1}d\right)^{-1}\rightarrow0$ and $a_{1}\rightarrow\infty$,
the relative energy level can be written as $E_{rel}=\left(2n+5/2\right)\hbar\omega$,
which has exactly the same form as the non-interacting spectrum. However,
there is a constant shift due to strong attractions. As a result,
the \emph{second} interacting energy level lines up with the \emph{first}
non-interacting energy level and so on. Their contributions to the
second virial coefficient cancel exactly with each other. In the end,
only the lowest energy level contributes to the coefficient. We thus
obtain immediately, 
\begin{equation}
\Delta b_{2,p}=\frac{1}{2}\left(2l+1\right)=\frac{3}{2},\label{eq:virial18}
\end{equation}
which is universal and temperature independent. The factor $2l+1$
accounts for the spectrum degeneracy. 

Near resonance with a small interaction range, we have $q\ll1$. Due
to the small size of $q$, to a good approximation, we can use $\exp\left[-\tilde{\omega}\left(2n+5/2+2q\right)\right]\simeq\left(1+2q\tilde{\omega}\right)\exp\left[-\tilde{\omega}\left(2n+5/2\right)\right]$,
so that: 
\begin{gather*}
\sum_{n=0}^{\infty}e^{-2n\tilde{\omega}}\frac{\Gamma\left(n+5/2\right)}{\left(n+5/4\right)\Gamma\left(n+1\right)}\simeq\frac{\sqrt{\pi}}{4\sqrt{2}}\cdot\frac{1}{\tilde{\omega}^{3/2}},
\end{gather*}
 and 
\begin{gather*}
\sum_{n=0}^{\infty}e^{-2n\tilde{\omega}}\frac{\Gamma\left(n+5/2\right)}{\left(n+5/4\right)^{2}\Gamma\left(n+1\right)}\approx\frac{\sqrt{\pi}}{\sqrt{2}}\cdot\frac{1}{\tilde{\omega}^{1/2}}.
\end{gather*}
Then, the second virial coefficient near resonance is given by,
\begin{gather}
\Delta b_{2,p}=\frac{3}{2}-\frac{3}{2}\tilde{\omega}\frac{1}{r_{1}d}\left[\left(\frac{d}{a_{1}}\right)^{3}+8\cdot\frac{\Gamma\left(5/4\right)}{\Gamma\left(-1/4\right)}\right]\nonumber \\
+\frac{3}{2\sqrt{2\pi}}\tilde{T}^{1/2}\left(\frac{k_{F}}{r_{1}}\right)+\frac{6\sqrt{2}}{\sqrt{\pi}}\tilde{T}^{-1/2}\left(\frac{1}{k_{F}a_{1}}\right)^{3}\left(\frac{k_{F}}{r_{1}}\right)^{2}.\label{eq:virial20}
\end{gather}

For the 3D \emph{anisotropic} case, it is difficult to derive the
asymptotic equation for the second virial coefficient near resonance
analytically. We have to solve the energy levels for $m=0$ and $m=\pm1$
respectively using Eqs. (\ref{eq:twobody15}) and (\ref{eq:twobody16}),
and calculate the second virial coefficient numerically. In Fig. 5
we calculate $\Delta b_{2,p}$ as a function of the dimensionless
interaction parameter $1/\left(k_{F}a_{1}\right)^{3}$ at three different
temperatures and a fixed small finite range of potential, $R=0.05d$.
Here we consider a gas with $N=10^{4}$ atoms and use the Fermi temperature
as the unit for temperature. All the curves with different temperatures
appear to cross at $1/\left(k_{F}a_{1}\right)\rightarrow0$. This
is the manifestation of universal behavior anticipated if the finite-range
corrections are small, meaning that there is no intrinsic length scale. 

\begin{figure}
\includegraphics[width=0.5\textwidth]{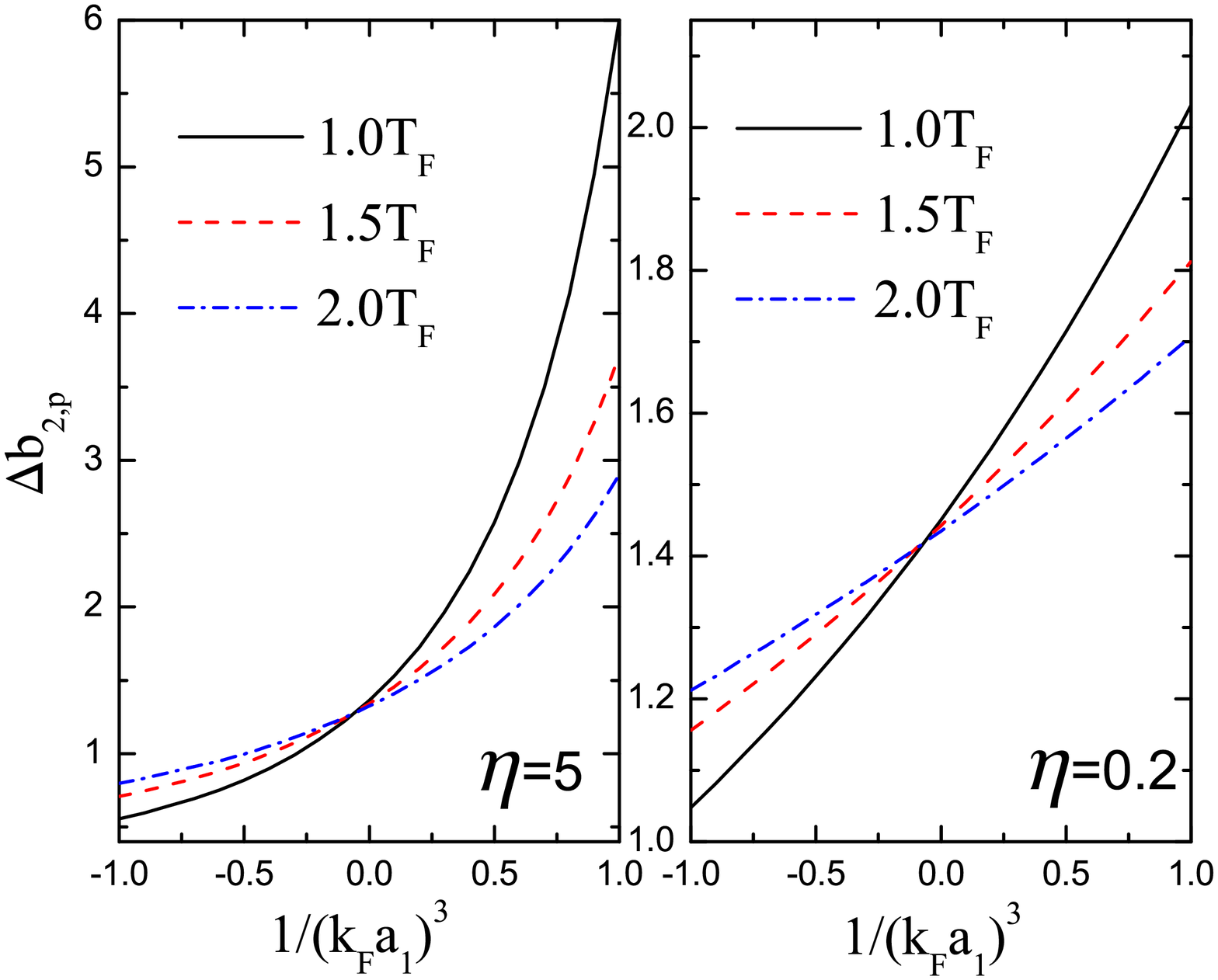}

\caption{(Color online) The second virial coefficient $\Delta b_{2,p}^{att}$
as a function of $1/\left(k_{F}a_{1}\right)^{3}$ with finite range
potential $R=0.05d$ in cigar-shaped trap $\eta=5$ and pancake-shaped
trap $\eta=0.2$. }

\label{Flo:fig5}
\end{figure}

In Fig. 6 we show the second virial coefficient $\Delta b_{2,p}$
at unitarity limit $a_{1}\rightarrow\pm\infty$ as a function of dimensionless
finite range of potential $R/d$. In the zero-range limit $\Delta b_{2,p}$
approaches the universal value of $3/2$ as $\tilde{\omega}=\left(3N\right)^{-1/3}\approx0.007$
is fairly small.

\begin{figure}
\includegraphics[width=0.5\textwidth]{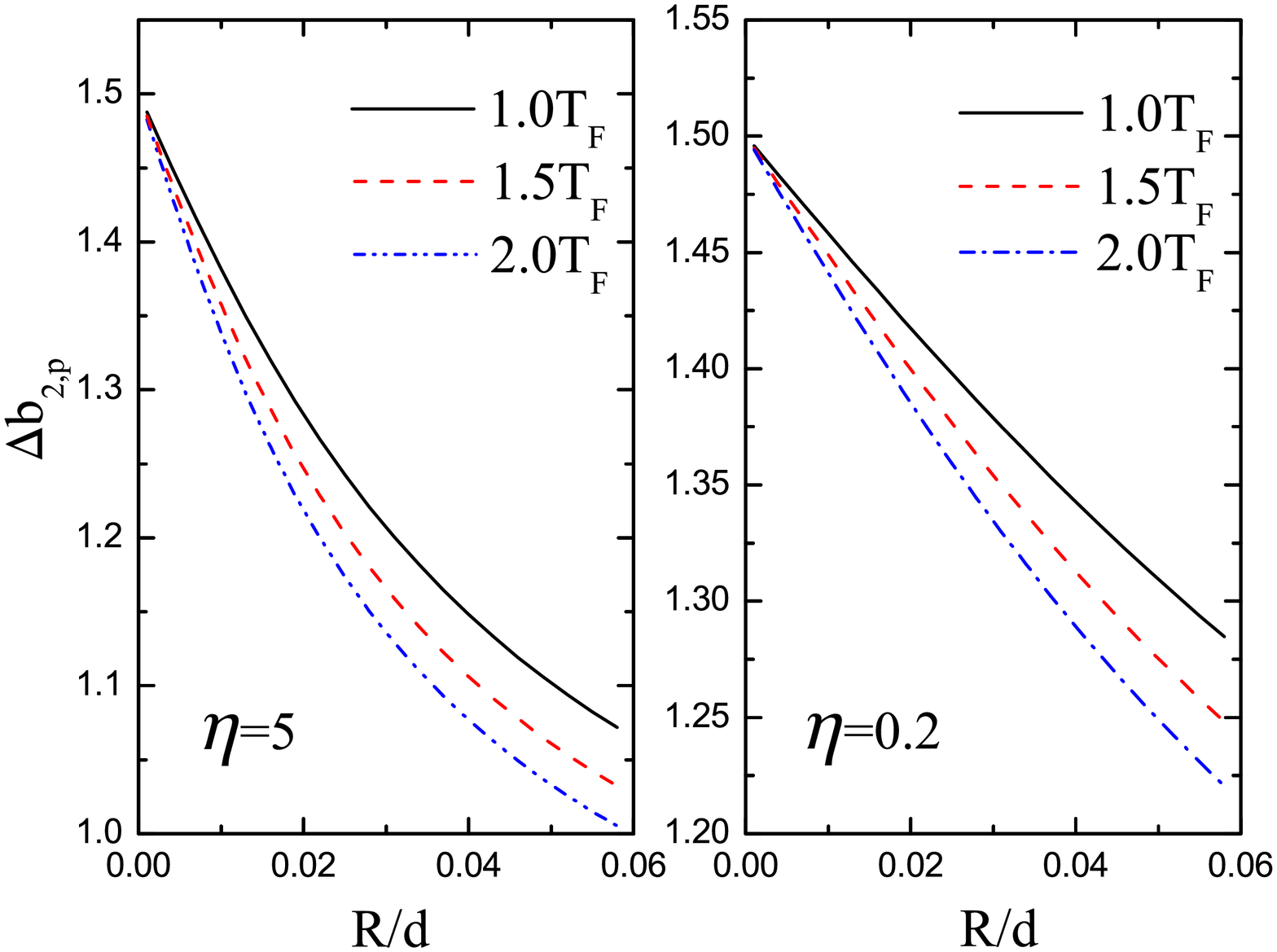}

\caption{(Color online) The second virial coefficient $\Delta b_{2,p}$ as
a function of the dimensionless finite range $R/d$ in the unitarity
limit in a cigar-shaped trap ($\eta=5$) and in a pancake-shaped trap
($\eta=0.2$).}

\label{Flo:fig7}
\end{figure}

\section{High-$T$ thermodynamics of strongly correlated fermions}

We are now in position to study the equation of state in the high-temperature
regime. Using the thermodynamic relation, we have the number equation,
\begin{gather}
N=N^{(1)}+2\frac{\left(k_{B}T\right)^{4}}{\left(\hbar\omega\right)^{3}}\left(2\Delta b_{2}z^{2}+3\Delta b_{3}z^{3}+\cdots\right),\label{eq:thermoN}
\end{gather}
and the total energy, 
\begin{gather}
E=-2\Omega+2\frac{\left(k_{B}T\right)^{4}}{\left(\hbar\omega\right)^{3}}\frac{T}{T_{F}}\left(\Delta b_{2}^{'}z^{2}+\Delta b_{3}^{'}z^{3}+\cdots\right),\label{eq:thermoE}
\end{gather}
where $\Delta b_{n}^{'}\equiv d\Delta b_{n}/d\tilde{T}$ and the non-interacting
number, 
\begin{gather}
N^{(1)}=-\frac{2\left(k_{B}T\right)^{3}}{\left(\hbar\bar{\omega}\right)^{3}}\frac{1}{2}\int\limits _{0}^{\infty}t^{2}\frac{ze^{-t}}{1+ze^{-t}}dt.\label{eq:thermoN1}
\end{gather}
The entropy of the system can be calculated using, 
\begin{gather}
S=\left(E-\Omega-\mu N\right)/T,\label{eq:entropy}
\end{gather}
where the chemical potential $\mu=k_{B}T\ln z$ (not to be confused
with the reduced mass). Eqs. (\ref{non_interacting_Omega_Trap}),
(\ref{eq:thermoN}), (\ref{eq:thermoE}), together with (\ref{eq:thermoN1}),
form a closed set of expressions for thermodynamics, which can be
solved self-consistently. 

Up to the second order virial expansion, the energy and entropy of
a Fermi gas with resonant $s$-wave and $p$-wave interactions are
shown in Figs. 7 and 8, respectively. Here, we consider a zero-range
potential so the thermodynamics is universal. The Fermi energy is
$E_{F}=k_{B}T_{F}=\left(3N\right)^{1/3}\hbar\omega$. For comparison,
we show the ideal gas result using the solid lines. It is clear that
the equation of state are strongly affected by interactions, even
in the high temperature regime. This interaction effect is particularly
apparent for $p$-wave interactions due to its large second virial
coefficient.

\begin{figure}
\includegraphics[width=0.5\textwidth]{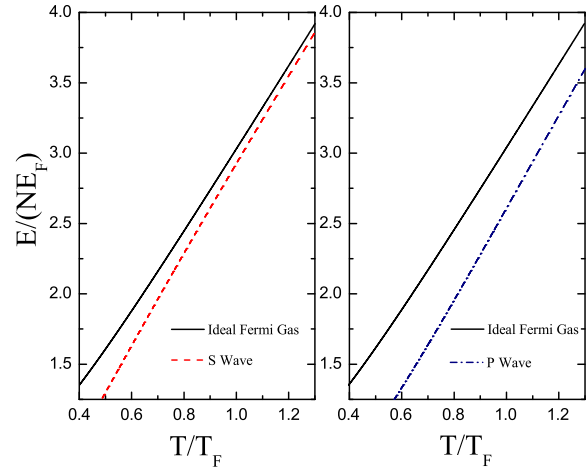}

\caption{(Color online) Temperature dependence of the energy of a strongly
correlated Fermi gas with $s$-wave and $p$-wave interactions. For
comparison, we plot the ideal gas result using solid lines.}

\label{Flo:fig8}
\end{figure}

\begin{figure}
\includegraphics[width=0.5\textwidth]{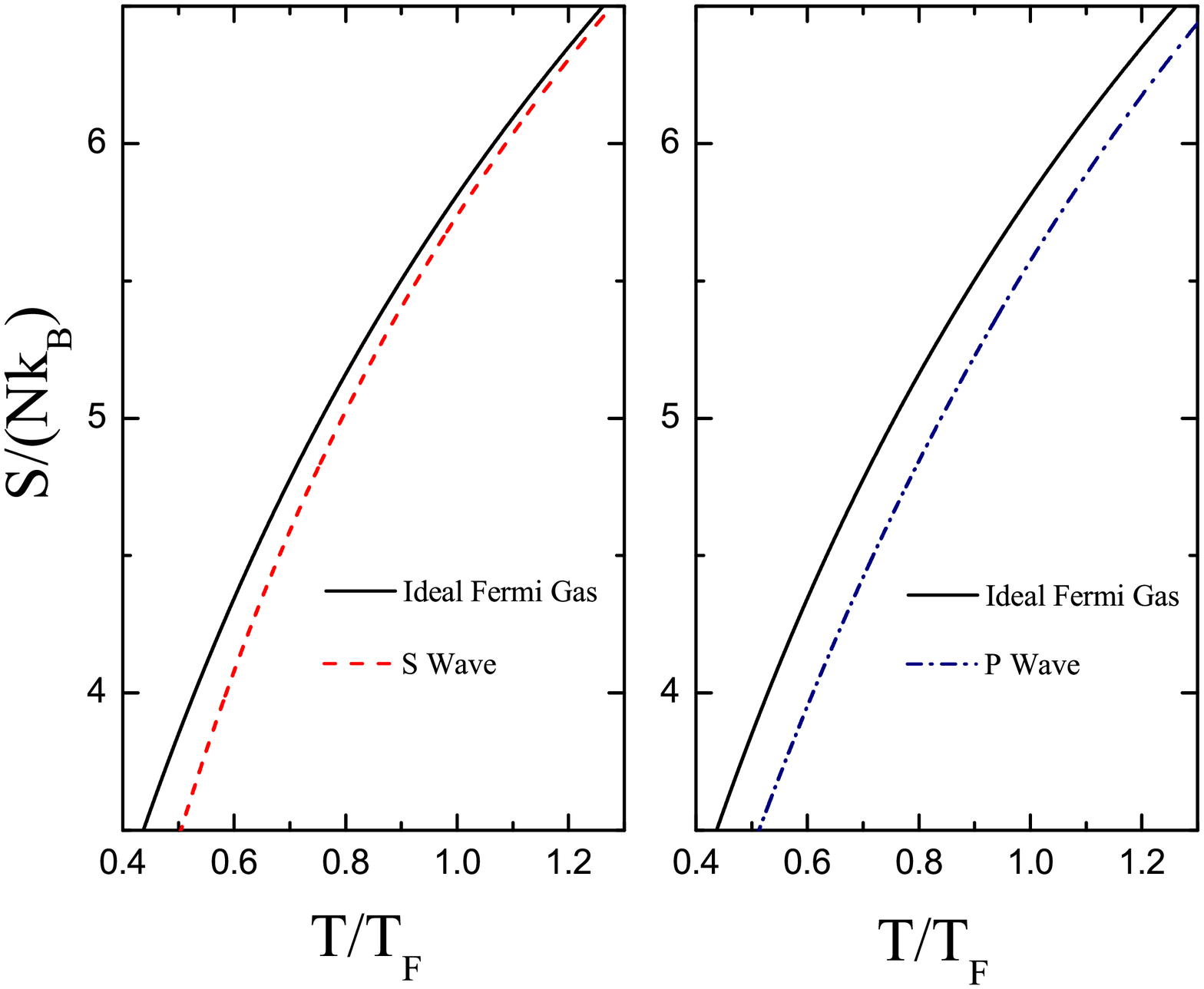}

\caption{(Color online) Temperature dependence of the entropy of a strongly
correlated Fermi gas with $s$-wave and $p$-wave interactions. The
ideal gas results are shown by the solid lines.}

\label{Flo:fig9}
\end{figure}

In Figs. 9 and 10, we present the non-universal effect on thermodynamics
caused by the presence of a finite-range of interaction potentials.
Eqs. (\ref{eq:5.10}) and (\ref{eq:virial20}) are used in the calculations
of the virial coefficient with the inclusion of the influence of the
finite-range of interactions.

\begin{figure}
\includegraphics[width=0.5\textwidth]{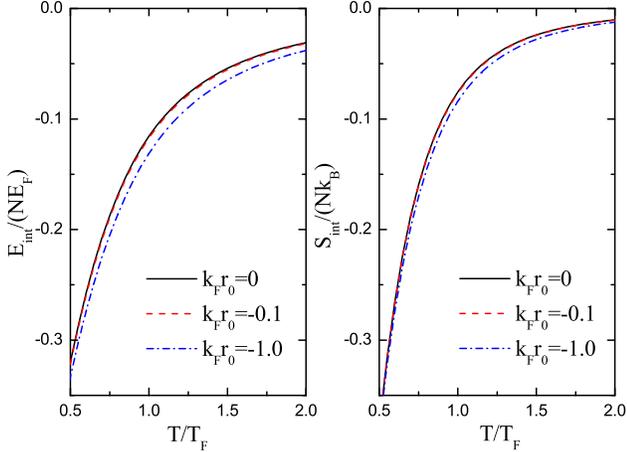}

\caption{(Color online) The non-universal effect on the energy and entropy
of a unitary $s$-wave Fermi gas. The non-universality arises from
the use of a finite-range of interaction potentials.}

\label{Flo:fig10}
\end{figure}

\begin{figure}
\includegraphics[width=0.5\textwidth]{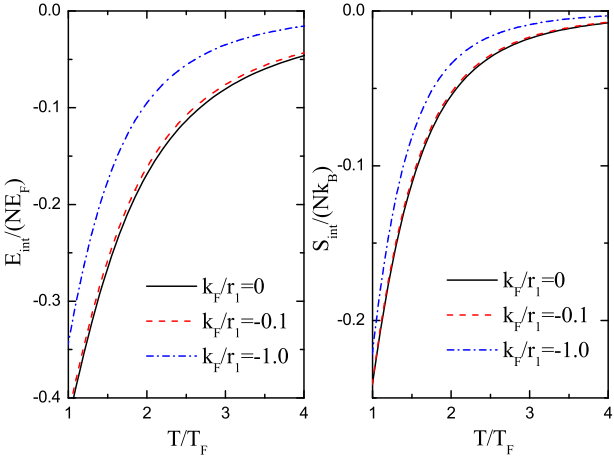}

\caption{(Color online) The same as in Fig. 9, but for a $p$-wave resonant
Fermi gas.}

\label{Flo:fig11}
\end{figure}

\section{Conclusions and remarks}

In conclusion, we have investigated the high-temperature thermodynamics
of a strongly correlated Fermi gas with $s$-wave and $p$-wave interactions,
by applying a second-order quantum virial expansion method. The second
virial coefficient has been calculated, based on the numerically exact
solution for two fermions in a harmonic trap. In this study, we have
particularly focused on the effects arising from a finite-range interaction
potential, which is crucial for $p$-wave interactions near the unitarity
limit. Non-universal corrections due to the finite range corrections
to the thermodynamics have been addressed. 

We expect that these thermodynamic results should be useful as a complementary
approach to quantum Monte Carlo simulations in understanding experimental
thermodynamic measurements for a $p$-wave Fermi gas near a Feshbach
resonance.
\begin{acknowledgments}
We acknowledge helpful discussions with Peng Zou and Hui Hu. This
work was supported by the ARC Discovery Projects DP0984637, DP0880404
and NFRPC Grant No. 2006CB921404. \end{acknowledgments}

\end{document}